%% file: sigconf.tex
\documentclass[sigconf]{acmart}
\AtBeginDocument{%
  \providecommand\BibTeX{{%
    \normalfont B\kern-0.5em{\scshape i\kern-0.25em b}\kern-0.8em\TeX}}}

\setcopyright{acmcopyright}
\copyrightyear{2021}
\acmYear{2021}

\acmConference[SIGIR '21]{SIGIR '21: ACM Special Interest Group on Information Retrieval}{July, 2021}{Montreal, Canada}

\acmSubmissionID{1358}

\usepackage{balance}
\usepackage{todonotes}
\usepackage{subcaption}
\begin{document}

\title{Evaluation of Field-Aware Neural Ranking Models for Recipe Search}


\author{Kentaro Takiguchi}
\email{gm19099@bristol.ac.uk}
\affiliation{%
  \institution{University of Bristol}
  \country{United Kingdom}
}

\author{Mikhail Fain}
\email{mikhail-fain@cookpad.com}
\affiliation{%
  \institution{Cookpad Ltd}
  \country{United Kingdom}}

\author{Niall Twomey}
\email{niall-twomey@cookpad.com}
\affiliation{%
  \institution{Cookpad Ltd}
  \country{United Kingdom}}

\author{Luis M Vaquero}
\email{luis.vaquero@bristol.ac.uk}
\affiliation{%
  \institution{University of Bristol}
  \country{United Kingdom}
}


\input{sections/01_abstract}

\ccsdesc[500]{Applied computing~Document management and text processing}
\ccsdesc[500]{Information systems~Learning to rank}

\keywords{recipes, field interactions, query-field, neural document ranking}

\maketitle

\input{sections/02_introduction_nt}
\input{sections/04_methodology_nt}

\input{sections/05_experiments_nt}

\input{sections/06_discussion}
\input{sections/07_conclusions}


\balance
\bibliographystyle{ACM-Reference-Format}
\bibliography{references}

\end{document}

%% file: sections/01_abstract.tex
\begin{abstract}

Explicitly modelling field interactions and correlations in complex document structures has recently gained popularity in neural document embedding and retrieval tasks. Although this requires the specification of bespoke task-dependent models, encouraging empirical results are beginning to emerge. We present the first in-depth analyses of non-linear multi-field interaction (NL-MFI) ranking in the cooking domain in this work. Our results show that field-weighted factorisation machines models provide a statistically significant improvement over baselines in recipe retrieval tasks. Additionally, we show that sparsely capturing subsets of field interactions based on domain knowledge and feature selection heuristics offers significant advantages over baselines and exhaustive alternatives. Although field-interaction aware models are more elaborate from an architectural basis, they are often more data-efficient in optimisation and are better suited for explainability due to mirrored document and model factorisation. 

\end{abstract}

%% file: sections/02_introduction_nt.tex
\section{Introduction}


In information retrieval, single-field document ranking has been a central topic since the early days of the field. However, it is typical that documents to be retrieved have multiple fields in practical applications. Recipes are a canonical example of multi-field data, consisting of titles, ingredient lists, procedures, and images.

In such applications, it is common to collapse complex, multi-field documents uniformly into embeddings \cite{MitraandCraswellNIR18}. This approach has been highly successful \cite{zhou2008large, pazzani2007content, lops2011content, balabanovic1997fab, logesh2019exploring, vairale2021recommendation}, but it is widely accepted that these data pipelines do not adequately capture how users view and interact with documents, nor do the models account for cross-field correlations \cite{pan2018field}.

Since the document's concept is the backbone that ties all fields together, correlated features will almost certainly be prevalent across fields. Feature duplication/overlap (resulting from correlated features) can result in more challenging loss surfaces and models are more likely to converge to local optima \cite{farrar1967multicollinearity, tolocsi2011classification}. 

Thus, de-correlating field representations empowers the model to learn field relevance from the training data and it further allows field weighting to be driven dynamically from context. 

Many model-agnostic approaches for producing field-aware document representations exist such as BM25F, an heuristic extension of BM25~\cite{Robertson2004}, Bayesian networks~\cite{piwowarski2003machine}, LambdaBM25~\cite{svore2009machine}, language modeling framework~\cite{Ogilvie03known}, probabilistic models~\cite{kim09prob}, or feedback weighted field relevance~\cite{kim2012field}.

There are also model-based approaches that learn query-field interactions such as Neural Ranking Models for Fields (NRM-F) \cite{DBLP:journals/corr/abs-1711-09174}, but fail to distinctly consider field-to-field interactions or account for correlated features. Other type of models could be extended to understand the value of complex field interaction modelling like field-weighted Factorization Machines (FwFM), which explicitly model feature interactions as arbitrary (non-linear) functions~\cite{pan2018field}.


We explore a field-aware modelling approach in the domain of a recipe search application, including a breadth of options for model architecture and construction to determine whether dynamic field weighting can reduce relevance dilution effects arising from the consideration of irrelevant fields in recipe search \cite{twomey2020towards}. 

Cooking has always been a vital daily routine for hundreds of millions of people, and under pandemic restrictions, it has brought family cohesion and mental health support for many people globally \footnote{\url{https://medium.com/cookpadteam/the-changing-face-of-italian-cooking-during-lockdown-7b1bbbcb2b56}}. Thus, improving retrieval tasks in the cooking domain has the potential to impart a positive impact on users of recipe recommendation services. 

Our models are trained on recipe, click, and query data from Cookpad's search platform; we evaluated these as a suite of Non-Linear Multi-Field Interaction (NL-MFI) configurations. 


The remainder of this document is structured as follows. Section 2 introduces our methodology, datasets, models, and evaluation protocol. Our experimental results and hypothesis evaluations are presented in Section 3, and we wrap up with discussions and conclusions in Section 4.


%% file: sections/04_methodology_nt.tex
\section{Methodology}

\subsection{Data and Preprocessing}
\label{sec:data_preprocessing}

This work focuses on recipe ranking. Recipes are complex multi-modal documents with several fields. We leverage existing search log data from the world's largest recipe community web service (\url{www.cookpad.com}) to learn and evaluate our models. 

Since our setting is recipe search, a key field to consider is the user's search query. Once a query is executed, set of candidate recipes are selected and served to a user based on the query text. The objective of this paper is to produce field-aware ranking models that are capable of improving the search experience of users. 

Supplementary to the query field, five additional recipe fields are used in our modelling: title, ingredients, description, country, and image. Most of these fields are textual and have variable size. The title field is usually short (consisting of $\approx 3.5$ words on average), whereas ingredients and descriptions are typically much longer consisting of multiple free-text fields. Recipes are often accompanied by a set of images (one main image for the recipe, and at most 3 per step). We incorporate image embeddings from the main recipe image (when available) since it captures global visual recipe traits and aesthetics.

It is clear that these fields are correlated. For example, given a recipe titled `pepperoni pizza', an experienced cook is likely to have prior expectations on the ingredients, steps and the look of the recipe.  Users interact with search in a variety of ways. For users that search with simple queries, like `chicken korma', one would expect that matches based on the recipe title is adequate. However, a wide variation of queries occur and users often with to filter by tool (`korma with a crock pot'), questions (`how can I bake cake without oven?'), specific sets of potentially negated ingredients (`cake without milk'), diets (`gluten free bread'), religious dietary restrictions (`halal chicken curry'), and other meta terms (`healthy/fast/easy soup') at different levels of specificity (`dinner' \textit{vs.} `sugar-free granola for breakfast'). This variation of query types demonstrates the organic interactions between users and search and the nuanced link between the query and user expectations. The opportunity to avoid prescribing actions based on query types but instead to learn behavioural patterns from interaction data is a key motivator for this research. 


We can completely reconstruct search sessions from the data logs. Each session is identifiable by a unique session ID, and aligned metadata is also available including event time, retrieved recipe IDs, clicked recipe ID, and query string. Over 99\% of queries are `clean' in the sense that the query is specifically targeting ingredients, dishes, meals etc. Other queries (e.g. `what is quinoa?') are removed when possible. We truncate the candidate recipe data at the last clicked position. Since we are aiming to optimise for click-through metrics, we implicitly treat the items which were examined by a user but not clicked as negatives. Only clicked items are positives and treated as the ground truth for training and evaluation.

We follow standard approaches for embedding images and text. We use average word embeddings for each field which, despite its simplicity, has been shown to achieve similar accuracy than recurrent units with significantly less training time \cite{DBLP:journals/corr/abs-1907-00937}. Image embeddings are obtained from pre-trained on ImageNet \cite{imagenet_cvpr09}.

\begin{figure}[t]
     \centering
     \begin{subfigure}[b]{0.99\linewidth}
         \centering
         \includegraphics[width=\textwidth]{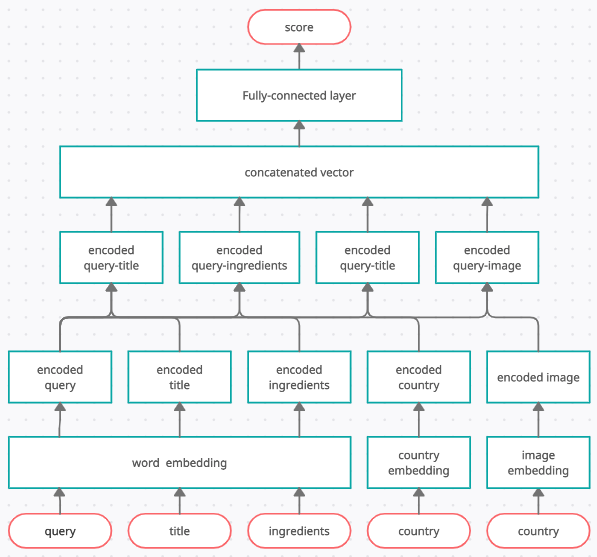}
         \caption{NRM-F architecture.}
         \label{fig:diagram_of_nrmf}
     \end{subfigure}\\
     \begin{subfigure}[b]{0.99\linewidth}
         \centering
         \includegraphics[width=\textwidth]{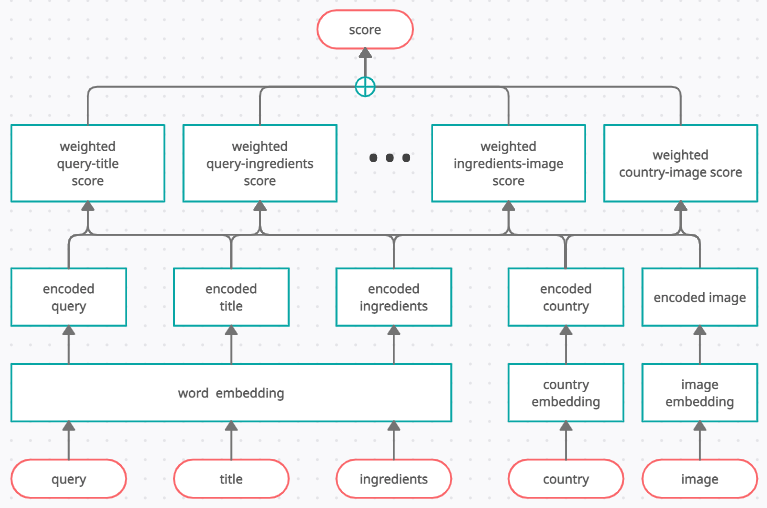}
         \caption{FwFM architecture}
         \label{fig:diagram_of_fwfm}
     \end{subfigure}
        \caption{Architecture of field-aware models considered.}
        \label{fig:architectures}
\end{figure}

\subsection{Models}
\label{sec:models}
We focus on the two field-aware models: NRM-F~\cite{DBLP:journals/corr/abs-1907-00937} and FwFM~\cite{pan2018field}. Our main objective is to assess their performance on recipe retrieval tasks. We will also investigate whether incorporating small architectural adjustments can improve retrieval performance.

\begin{table}[H]
    \centering
    \caption{Model and field interaction configurations of the two architectures as specified in original publications.}
    \begin{tabular}{lll}
        \toprule
        & NRM-F & FwFM \\
        \midrule
        First-order features & Not used & Used \\
        Interaction selection & Query-field & All \\
        Interaction representation & Hadamard product & Dot product \\
        Interaction aggregation & Concatenation & Weighted sum \\
        \bottomrule
    \end{tabular}
    \label{tab:differences_in_models}
\end{table}

The architecture of these two models is shown in Figure \ref{fig:architectures} and the key differences are summarised by interactions, representations and aggregations (see Table \ref{tab:differences_in_models}).
%
Field interactions are broadly categorised 2-nd order interactions (consisting of $k(k-1)/2$ pairwise combinations across fields) and 1-st order fields (consisting of $k$ individual fields). The original NRM-F model specifies `query-field' interactions, where the `query' field is paired with all other 1-st order interactions (yielding $k-1$ interaction pairs). We also use the term `all' interactions, which refers to the union of 1-st and 2-nd order interactions, and this was used by FwFM (giving $k(k+1)/2$ interaction pairs). 

\subsection{Evaluation}

We use Normalised Discounted Cumulative Gain (NDCG) at 20 to evaluate models, which measures the quality of the first page of results. We evaluate performance with multiple pairwise t-tests with the Bonferroni correction with a (fairly stringent) threshold of significance set to $\alpha=0.01$. Performance is evaluated with 10-fold cross validation for parameter selection. Each train fold contains about 33,000 search sessions on average.

All code for the proposed work is available on GitHub \footnote{\url{https://github.com/rejasupotaro/field-interactions-in-document-ranking}}.

%% file: sections/05_experiments_nt.tex
\section{Experiments}

Our experiments explore variants of the FwFM and NRM-F models on recipe retrieval based on interaction pairings (`1-st', `2-nd', `query-field', and `all' which are defined in Section \ref{sec:models}). 

\subsection{Effect of Interaction Selection}
\label{subsec:q2ff2f}

In this section, we investigate the effect of various interaction selection mechanisms on both models. Specifically we consider `query-field' and `all' interactions (defined in Section \ref{sec:models}) on FwFM and NRM-F models. Intuitively, the `query-field' interactions result in simpler models since fewer cross-field interactions are derived, but the `all' configuration may be better able to learn complex interactions. As a baseline we consider a basic model which is the concatenation of all 1-st order features, and we also report performance of random models. 

NDCG@20 scores over the five model configurations are shown in Table \ref{tab:rq2_result}. We can note that the query-field interactions tended to consistently out-perform the others on both NRM-F and FwFM. Noting that this configuration is a subset of the `all' configuration, it is interesting to see that the `simpler' model out-performed the more complete interaction specification.  The best performing model was `FwFM (query-field)', whose results were significantly better than all others considered. 
We note also that we explored this model configuration to understand which configuration is most optimal for recipe re-ranking tasks. This novel configuration was not explored in the original FwFM paper. Random performance is relatively high because models are evaluated on search sessions that are cut-off at the last clicked position.


\begin{table}
    \centering
    \small
    \caption{Comparison of NRM-F and FwFM models on mean NDCG scores. The FwFM query-field results are statistically significant better than all other configurations.}
    \begin{tabular}{ll}
        \toprule
        Model & NDCG@20 \\
        \midrule
        Random & 0.523 \\
        Baseline & 0.643 \\
        NRM-F model (all) & 0.641 \\
        NRM-F model (query-field) & 0.652 \\
        FwFM model (all) &  0.661 \\
        FwFM model (query-field) & 0.667 $^\dagger$ \\
        \bottomrule
    \end{tabular}
    \label{tab:rq2_result}
\end{table}

\begin{figure}[h]
    \centering
    \includegraphics[width=0.9\linewidth]{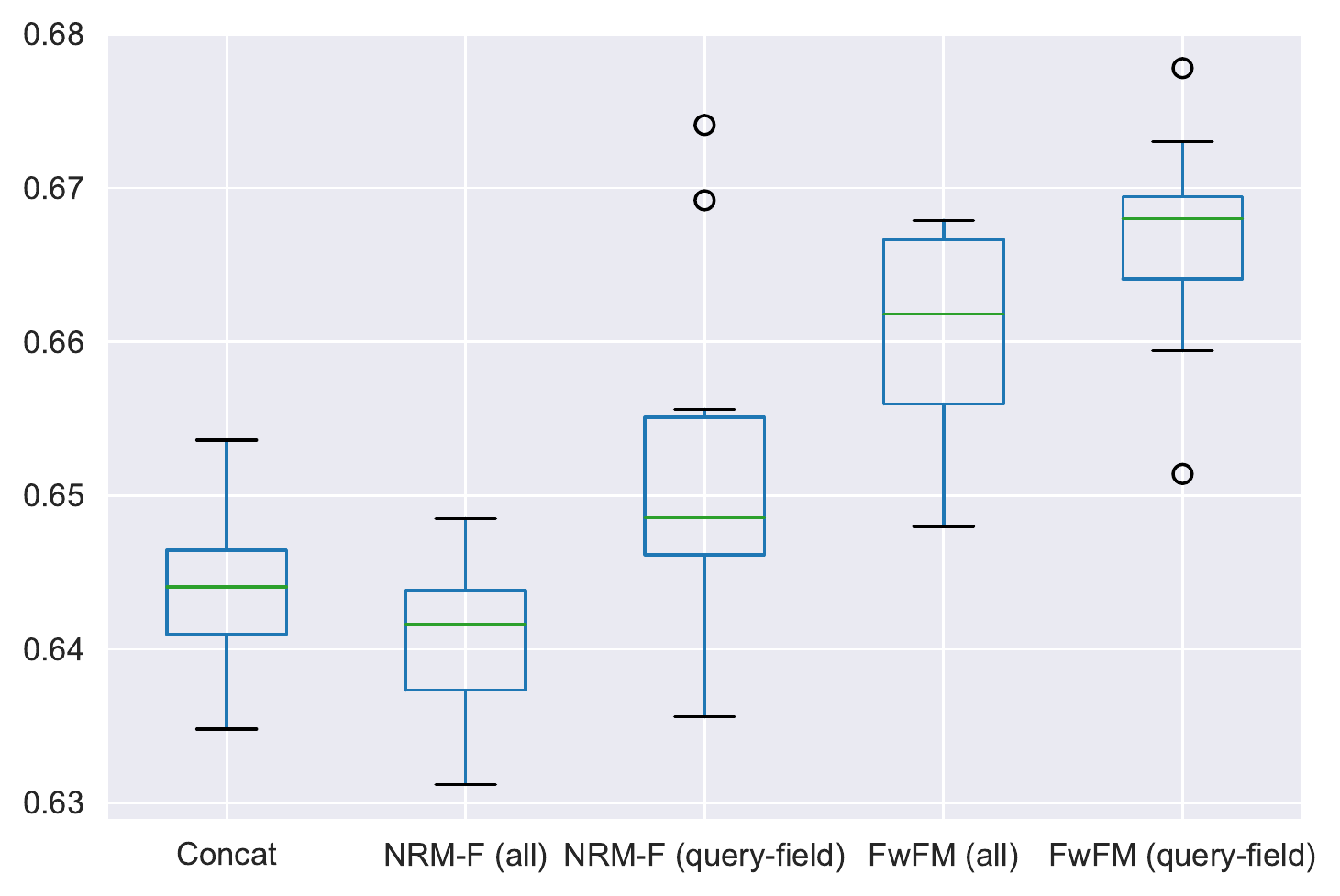}
    \caption{This image shows the spread of NDCG scores for various model configurations. We can see that FW query-field results are more tightly packed.}
    \label{fig:rq2_boxplot}
\end{figure}

The distribution of results are shown in Figure \ref{fig:rq2_boxplot}. We can see that as well as having a higher median, the spread and inter-quartile range of this configuration is narrower. 

\subsection{Non-Linear Field Interactions}

We continue the exploration on new interaction pair configurations in this section, specifically to understand the value given by 1-st order interactions. Table \ref{tab:1st2nd} shows the average NDCG scores for the above models. The results for the FwFM have not significantly changed, though they dipped slightly. However, the average results for the NRM-F based model have improved significantly over the baselines presented in Section \ref{subsec:q2ff2f}.
FwFM models are still the top-performing on this dataset, and `FwFM 1-st $\&$ 2-nd' remains significantly better than the rest.


\begin{table}
    \centering
    \caption{Comparison of models with different feature sets. FwFM with 1-st and 2-nd interactions are statistically significant better than all other configurations.
    }
    \begin{tabular}{llll}
        \toprule
        Model & 1-st & 2-nd & NDCG@20 \\
        \midrule
        NRM-F &            & \checkmark & 0.652 \\
        NRM-F & \checkmark & \checkmark & 0.650 \\
        FwFM  &            & \checkmark & 0.645 \\
        FwFM  & \checkmark & \checkmark & 0.661 $^\dagger$ \\
        \bottomrule
    \end{tabular}
    \label{tab:1st2nd}
\end{table}

\subsection{Exploring High-Yield Interactions}

In this section we take an ablation approach to understand whether the inclusion of novel interaction pairs can improve on the query-field results. The correlation between the individual interaction scores and ground truth click labels from the FwFM model trained on all 1-st and 2-nd order interactions is calculated. FwFM was chosen because it is the strongest performing model on our dataset. This correlation score is assumed to act as a proxy indicator to field importance, and the interactions are sorted by their correlation score in order to help us understand the pairs which are deemed important. 

\begin{figure}
    \centering
    \includegraphics[width=0.9\linewidth]{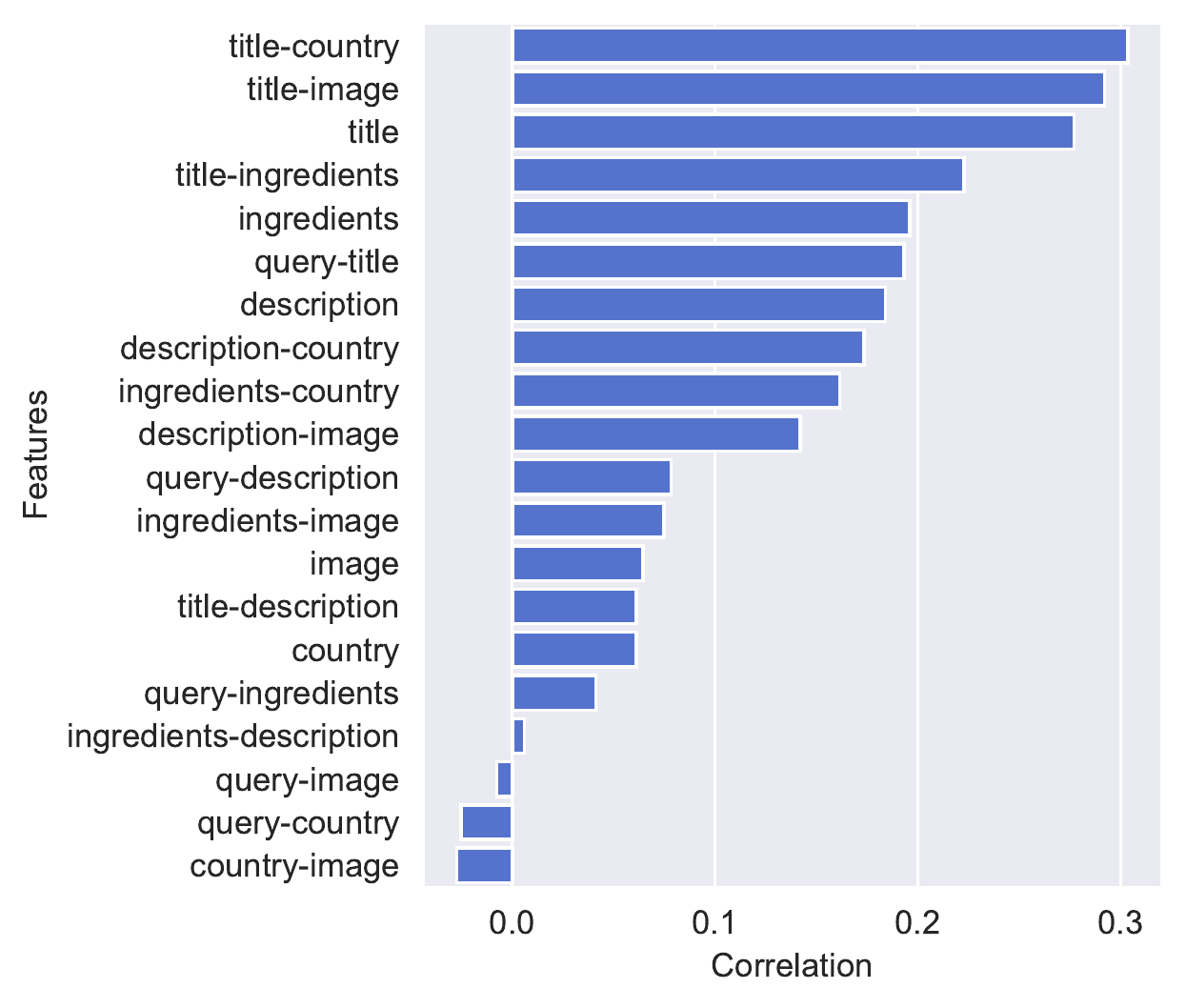}
    \caption{This figures shows the calculated correlation between interaction terms and the click labels. Larger values indicate closer connections.}
    \label{fig:rq4_corr}
\end{figure}

Figure \ref{fig:rq4_corr} shows the correlations between interaction and click labels. 
\textit{A priori}, one may have expected query-field interactions to be the most highly correlated to labels in search ranking tasks. However, a simple inspection of Figure \ref{fig:rq4_corr} shows that `query' fields are not highly ranked. To understand this, we must remember that our ranking task follows an initial candidate generation phase which we do not influence (see Section \ref{sec:data_preprocessing}). As a result the items that we score are implicitly dependent on and correlated to the query even when not directly including the query. A complementary perspective is that the query-field interactions do not add significant \textit{new} information to the scoring model, implying that the interactions that are most highly correlated to the labels are those which contribute new information. We note too that the query-field set of interactions would almost certainly be the most highly scored fields in candidate generation tasks. 

Interactions on other field pairs build on the implicit query dependency of candidates and introduce new features which lead to them being highly ranked. The highest pair is `title-country.' We understand the importance of this arising due to regional preferences being strong indicators of interest in and of itself. 
We were surprised to see that image features have relatively low weight. This is a valuable insight of the experiment since \textit{a priori} image aesthetics were assumed would contribute highly to clicks due to their visual appeal. The low correlation on all image pairs indicates that image features (which are expensive to compute) may be a candidate for removal. 

We leverage these results and construct a correlation-driven interaction model on a sparse subset of interactions. Interaction pairs are selected from the validation set. In Table \ref{tab:rq4_result} we can see that we can improve over baseline performance with this approach. Both `query-field' and `selected' offer a statistically significant improvement over `all' but with many fewer interactions modelled. The performance of `query-field' and `selected' configurations is similar, with no significant differences between them. Although we have not beaten the query-field results with correlation selection, we believe this approach is a valuable contribution since it extends the scope for these models with improved performance. For example, query-field interactions cannot be used in re-ranking recipe feeds since no `query' field exists, and performance in these situations may benefit from restricted interactions achieved by the correlation-driven interaction selection.

\begin{table}
    \centering
    \caption{The results for the interaction-oriented experiments. The performance of `FwFM (all)' is significantly worse than the other two models.}
    \begin{tabular}{ll}
        \toprule
        Model & NDCG@20 \\
        \midrule
        FwFM (all) & 0.661 \\
        FwFM (selected) & 0.663 $^\dagger$ \\
        FwFM (query-field) & 0.667 $^\dagger$\\
        \bottomrule
    \end{tabular}
    \label{tab:rq4_result}
\end{table}

%% file: sections/07_conclusions.tex
\section{Conclusions}

Strong early evidence is provided in this paper for the broad advantages provided by non-linear multi-interaction field models in the domain of recipe ranking. By reducing between-field feature correlations and methodically calculating significant improvement over baselines, we  demonstrate how new insights can be gleaned for practitioners about their domain. We believe that we have shown that non-linear multi-field interaction models investigated are strong candidates for the domain. The next steps for this research are to explore the value of these models in online experiments and across different domains and tasks.


